\documentclass[onecolumn,prd,superscriptaddress,nofootinbib,notitlepage]{revtex4-1}
\usepackage{latexsym}
\usepackage{amsmath,amsfonts,graphicx,accents}
\usepackage{amsbsy}
\usepackage{hyperref}
\usepackage{mathrsfs}
\usepackage{color}
\usepackage{psfrag}
\usepackage{enumerate}
\usepackage{amsmath,amssymb,calc,amsfonts}
\usepackage{latexsym}
\usepackage[utf8]{inputenc}
\usepackage[percent]{overpic}
\usepackage[autostyle=false, style=english]{csquotes}
\MakeOuterQuote{"}
\DeclareFontFamily{U}{rsfs}{}         
\DeclareFontShape{U}{rsfs}{m}{n}{<5> rsfs5 <6><7> rsfs7          %
  <8><9><10><10.95><12><14.4><17.28><20.74><24.88> rsfs10}{}     %
\DeclareMathAlphabet{\mathfs}{U}{rsfs}{m}{n}                     %
\definecolor{indiagreen}{rgb}{0.07, 0.53, 0.03}
\def\beq{\begin{eqnarray}}
\def\eeq{\end{eqnarray}}

\def\={\stackrel{\Delta}{=}}

\begin{document}
\title{Topological Nature of Black Hole Solutions in dRGT Massive Gravity}

\author{C. Fairoos}\email{fairoos.phy@gmail.com}
\affiliation{T. K. M. College of Arts and Science Kollam-691005, India}
\author{T. Sharqui}\email{tsharqui0@gmail.com}
\affiliation{Department of Physics, \\
School of Physical, Chemical and Applied Sciences,\\
Pondicherry University, Puducherry-605014, India}

\begin{abstract}
We study the thermodynamic properties of black holes in dRGT massive gravity theory using Duan's $\phi-$ mapping topological current theory. The topological features and the corresponding thermodynamic stability conditions for neutral and charged cases are discussed. A neutral black hole in four dimensions has a topological number $0$, sharing the same topological class of $D\geq6$ Gauss-Bonnet-AdS black hole. We show that the charged black hole in four-dimensional massive gravity has the same topological structure as the AdS-RN black hole. Further, we have extended the calculations to higher dimensions. Our calculations strengthen the conjecture that the addition of  higher interaction terms to Einstein-Hilbert action does not alter the topological number of black holes in four spacetime dimensions. However, in higher dimensional massive gravity, the topological number indeed depends on the black hole parameters.
\end{abstract}

 \maketitle
\section{Introduction}
The study of black holes, especially their thermodynamic properties is a promising tool to understand the quantum nature of gravity. This idea is inspired by the correspondence between ordinary thermodynamics and statistical mechanics which suggests the black hole entropy \cite{Bekenstein:1972tm, Bekenstein:1973ur} should have a state-counting interpretation in terms of quantum degrees of freedom near the horizon. A remarkable discovery in this pursuit was the Hawking-Page phase transition \cite{Hawking:1982dh}, which provides a concrete analogy between thermodynamics and gravity. Hawking and Page found that there is a phase transition between a black hole and thermal radiation in Anti-de Sitter (AdS) spacetime. These transitions are then interpreted as the confinement/de-confinement phase transitions in the context of AdS/CFT correspondence\cite{Witten:1998qj}. Charged black holes in AdS space-time exhibit small-large phase transition analogous to the liquid-gas phase transition of van der Waal fluid \cite{Kubiznak:2012wp}. Subsequently, the quest for the underlying microstructure of black holes exploiting phase transitions has progressed by employing different approaches \cite{Wei:2019uqg, Wei:2019yvs}. One of the approaches to investigate the thermodynamic features of black holes in extended phase space is the topology. \\

The key point in the topological theory is that black holes can be treated as defects in the thermodynamic parameter space. In topological studies, the properties of a given field configuration $\phi(\vec{x})$ are deduced from the zero points of the field in space, i.e., $\phi(\vec{x})|_{\vec{x}=\vec{z}} =0$\cite{Wei:2022dzw}. One of the topological quantities associated with the zero point of a field is its winding number which can be obtained using Duan's $\phi-$ mapping topological current theory \cite{Duan:1979ucg, Duan:1984ws}. The winding number reflects the local thermodynamic stability of the system. Using winding numbers, one can deduce the topological number of the black hole solution that determines the global thermodynamic stability. In \cite{Wei:2022dzw}, Schwarzschild, Reissner-Nordstrom(RN), and RN-AdS black holes are considered. The topological nature is also discussed for Gauss-Bonnet black holes in AdS space\cite{Yerra:2022alz, Liu:2022aqt}, black holes in de Sitter \cite{Du:2023wwg}, and BTZ black holes \cite{Du:2023nkr}. In all cases, the black hole solutions are classified into three topology classes based on their topological number.\\

The general theory of relativity (GR) is a massless spin two-field theory that describes how gravity works in four space-time dimensions. The massive gravity theory was formulated as a straightforward modification of general relativity in which the graviton acquires a non-zero mass. The advantage of such a theory is that it explains the accelerated expansion of our universe without introducing a bare cosmological constant. The effect of putting mass to gravitons essentially modifies GR by weakening it at large scales, which allows the universe to accelerate. However, the  predictions of massive gravity theory at small scales are the same GR. The initial massive gravity model was proposed by Fierz and Pauli in 1939 \cite{Fierz:1939ix}. However, the theory did not produce correct GR limits in the massless case. Also, the non-linear modifications of the initial theory lead to "Boulware-Deser" ghost instability \cite{Boulware:1972yco}. Later, de Rham, Gabadadze, and Tolley (dRGT) proposed a  "Boulware-Deser" ghost-free theory which reduces to GR in the massless case\cite{Ghosh:2019eoo}.  Consequently, the black hole solutions in this modified gravity theory and their thermodynamic characteristics were extensively investigated. The van der Waal-like features and other related topics such as triple point, Reentrant phase transitions, heat engines, and throttling process were also studied \cite{Zou:2016sab, Liu:2018jld, Hendi:2017bys, Yerra:2020bfx}. In addition, several methods to probe the microstructure of black hole solutions in massive gravity were  presented using various thermodynamic-geometry approaches \cite{Chabab:2019mlu, Wu:2020fij, Yerra:2020oph, Safir:2022vkf, Safir:2023thg}.\\

This article aims to study the nature of thermodynamic stabilities of black hole solutions in dRGT massive gravity theory exploiting Duan's $\phi-$ mapping topological current theory. The structure is the following: In the next section, we briefly discuss the thermodynamic features of black holes in dRGT massive gravity theory. In \ref{topology}, we review the $\phi-$ mapping topological current theory as discussed in  \cite{Wei:2022dzw}. The topological nature of black hole solutions in four-dimensional dRGT massive gravity theory is presented in \ref{massive_topology}. The calculations are extended to five dimensions in \ref{massive_topology_2} and our observations are presented in \ref{dis}.
\section{Black Hole Thermodynamics in Higher Dimensional \lowercase{d}RGT Massive Gravity}\label{thermodynamics}

We begin with a brief discussion of black hole solutions in $(n+2)$-dimensional  dRGT massive gravity coupled to a non-linear electromagnetic field in AdS space described by the following action \cite{Cai:2014znn},
\begin{equation}\label{action}
S= \frac{1}{16 \pi} \int d^{n+2} x \sqrt{-g} \left[ R+\frac{n(n+1)}{L^2}+m^2\sum_{i=1}^4 c_i \ \mathcal{U}_i(g,f)-\frac{1}{4 } F_{\mu \nu}F^{\mu \nu}\right],
\end{equation} 
 where $F_{\mu \nu}=\partial_\mu A_\nu-\partial_\nu A_\mu $ is the electromagnetic field tensor with vector potential $A_\mu$, $L$ is the AdS radius which is related to the cosmological constant $\Lambda$ by ${n(n+1)}/({2L^2})=-\Lambda$. Further, $m$ is related to the graviton mass, $c_i$ are coupling parameters, and $f_{\mu \nu}$ is the reference metric coupled to the space-time metric $g_{\mu \nu}$. The symmetric polynomials  $\mathcal{U}_i$ are the graviton interaction terms and are obtained from a $(n+2) \times (n+2) $ matrix $\mathcal{K}^\mu_\nu =\sqrt{g^{\mu \alpha} f_{\nu \alpha}}$, given by,
\begin{eqnarray*}
\begin{split}
\mathcal{U}_1 &= [\mathcal{K}],\\
\mathcal{U}_2 & = [\mathcal{K}]^2-[\mathcal{K}^2],\\
\mathcal{U}_3 & = [\mathcal{K}]^3-3[\mathcal{K}^2][\mathcal{K}]+2[\mathcal{K}^3],\\
\mathcal{U}_4 & = [\mathcal{K}]^4-6[\mathcal{K}^2][\mathcal{K}]^2+8[\mathcal{K}^3][\mathcal{K}]+3[\mathcal{K}^2]^2-6[\mathcal{K}^4],
\end{split}
\end{eqnarray*}
where $\mathcal{K}_{\mu \nu}^2 = \mathcal{K}_{\mu \alpha} \mathcal{K}^\alpha_\nu$, and the rectangular brackets denote traces. Consider the metric function given by,

\begin{equation}\label{metric}
ds^2=-f(r) \ dt^2+\frac{1}{f(r)} \ dr^2+r^2 \ h_{ij} dx^i dx^j,
\end{equation}

where the line element $h_{ij} dx^i dx^j$ describes the Einstein space with a constant curvature of $n(n-1)k$. The topological parameter $k$ characterizes the geometry of the black hole horizon hypersurface. It takes values 0, -1, or 1, representing planar, hyperbolic, and spherical topology, respectively. A detailed analysis of black hole solutions with various horizon topologies is given in \citep{Cai:2014znn}. Taking $f_{\mu\nu}=diag(0,0,1, c_0^2 h_{ij})$, with $c_0$ being a positive constant, we get,

\begin{eqnarray*}
\mathcal{U}_1 &= &\frac{nc_0}{r},\\
\mathcal{U}_2 & = &n\left(n-1\right)\frac{c_0^2}{r^2},\\
\mathcal{U}_3 & =& n\left(n-1\right)\left(n-2\right) \frac{c_0^3}{r^3},\\
\mathcal{U}_4 & =& n\left(n-1\right)\left(n-2\right) \left(n-3\right)\frac{c_0^4}{r^4}.
\end{eqnarray*}

Note that the values of $\mathcal{U}_i$ depend on the spacetime dimesnsions. Now, the metric function $f(r)$ in Eq. (\ref{metric}) is given as \cite{Cai:2014znn},

\begin{eqnarray}\label{metric_fn}
f(r) &=& k - \frac{16 \pi M}{n V_n r^{n-1}}-\frac{2 \Lambda}{n(n+1) \pi} r^2 + \frac{Q^2}{2n(n-1) r^{2(n-1)}}
+ \frac{c_0 c_1 m^2}{n} r \\ \nonumber
&+& c_0^2 c_2 m^2 + \frac{(n-1) c_0^3 c_3 m^2}{r}+ \frac{(n-1)(n-2) c_0^4 c_4 m^2}{r^2},
\end{eqnarray}
here, $M$ and $Q$ are the black hole mass and charge, respectively. $V_n$ is the volume of space spanned by $\{x_i\}$. Note that in the limit $m\rightarrow0$, the metric function reduces to $(n+2)$ dimensional Schwarzschild black hole in AdS space. In the extended phase space, $\Lambda$ behaves like thermodynamic pressure with the following relation:
\begin{equation*}
P = -\frac{\Lambda}{8\pi}.
\end{equation*}
Now, the event horizon ($r_h$) is determined by the largest root of the equation $f(r)|_{r=r_h}=0$. The Hawking temperature of the black hole is related to its surface gravity by the relation $T_H = {\kappa}/({2\pi})$, where the surface gravity $\kappa = f'(r_h)/2$. The mass $M$, temperature $T_H$, and entropy $S$ are expressed in terms of the horizon radius  $r_h$ and pressure as following:

\begin{eqnarray}\label{m}
M &=& \frac{n V_n}{16 \pi} r_h^{n-1}\Big[ k+ \frac{16 \pi P}{n(n+1)} r_h^2 + \frac{Q^2}{2n(n-1) r_h^{2(n-1)}}
+ \frac{c_0 c_1 m^2}{n} r_h, \\ \nonumber
&+& c_0^2 c_2 m^2 + \frac{(n-1) c_0^3 c_3 m^2}{r_h}+ \frac{(n-1)(n-2) c_0^4 c_4 m^2}{r_h^2}\Big],
\end{eqnarray}
\begin{eqnarray}\label{t}
T_H &=& \frac{1}{4\pi r_h}\Big[ (n-1)k + \frac{16 \pi P}{n} r_h^2 - \frac{Q^2}{2 n r_h^{2(n-1)}}+c_0 c_1 m^2 r_h + (n-1) c_0^2 c_2 m^2\\ \nonumber
&+&  \frac{(n-1) (n-2) c_0^3 c_3 m^2}{r_h}+\frac{(n-1)(n-2)(n-3) c_0^4 c_4 m^2}{r_h^2}\Big],
\end{eqnarray}
\begin{equation}\label{s}
S = \frac{V_n}{4} r_h^n.
\end{equation}
The first law of black hole thermodynamics is readily obtained if we consider the black hole mass $M$ as the enthalpy of the gravitational system \cite{Xu:2015rfa}. To proceed further, we make certain simplifications. First, we will restrict ourselves to spherically symmetric solutions, i.e., $k=1$. Also, we set $c_0=1$ without the loss of generality.\\

Now, we discuss the black hole stability conditions in terms of the parameters of the theory for $n=2$. In four dimensions, the values of $\mathcal{U}_i$ are: $\mathcal{U}_1=\frac{2}{r}$,~~ $\mathcal{U}_2=\frac{2}{r^2}, and ~~\mathcal{U}_3= \mathcal{U}_4= 0 \ (c_3=c_4=0)$. The spherically symmetric black hole solution is described by the metric function give by Eq. (\ref{metric_fn}), 
\begin{equation}\label{metric_1}
f(r)=1-\frac{2M}{r}-\frac{\Lambda r^2}{3} +\frac{Q^2}{r^2}+m^2\left( \frac{ c_1}{2}r +c_2\right).
\end{equation}
Now, the corresponding expressions for $M$, $T$, and $S$ become Eq. (\ref{m}, \ref{t},\ref{s}),
\begin{align}
M=&\frac{r_h}{2}\left(1+\frac{8\pi}{3} P r_h^2 + \frac{Q^2}{r_h^2}+m^2 \left(\frac{c_1}{2} r_h + c_2\right)\right),\\
T_H=& \frac{1}{4 \pi r_h}\left( 1+8 \pi P r_h^2 -\frac{Q^2}{r_h^2}+ m^2\left(c_2+c_1 r_h\right)\right),
\end{align}
\begin{equation}
S = \pi r_h^2.
\end{equation}
One can analyze the behavior of black hole temperature as a function of horizon radius. The local minima and local maxima of black hole temperature are obtained from the condition,
\begin{equation}
\frac{\partial T_H}{\partial r_h}=0.
\end{equation}
Solving the above condition, we obtain the expressions for small and large black hole radii, indicating two black hole phases. Therefore, a positive local minimum of Hawking temperature indicates the first-order phase transition similar to a van der Waal system \cite{Safir:2023thg}. As the phase transition in black holes is caused by changing the horizon radius, it also changes the arrangements of energy and matter inside the hole, i.e., the characteristics of phase transition are directly linked to the micro-structure of black holes \cite{Wei:2015iwa}. An important aspect in this regard is the stability of the thermodynamic system. The global stability can be examined by looking at the free energy profile of the system \cite{Hawking:1982dh}. In the canonical ensemble approach, we consider the black hole as a closed system and is described by Helmholtz free energy,
\begin{eqnarray} \label{free}
\mathcal{F} &=& M- T_H S\\ \nonumber
&=& \frac{r_h}{4}\left(1- \frac{8\pi}{3} P r_h^2+\frac{3 Q^2}{r_h^2}+m^2 c_2\right).
\end{eqnarray}
Note that a negative $\mathcal{F} $ describes a stable thermodynamic system \cite{Cai:2013qga}. Therefore, black hole phase transition exists if $\mathcal{F}\le 0$, i.e.,
\begin{equation}
\frac{8 \pi}{3} P r_h^2 \leq 1+\frac{3 Q^2}{r_h^2}+ m^2c_2.
\end{equation}
A relevant quantity in thermodynamics is the heat capacity at constant pressure $C_P$ which reflects the local thermodynamic stabilities. A positive heat capacity corresponds to a locally stable thermodynamic system whereas a negative value of $C_P$ indicates local instability.
\begin{eqnarray}
C_P = T_H\left(\frac{\partial S}{\partial T_H}\right)_P = T_H \left( \frac{\partial_{r_h} S}{\partial_{r_h} T_H}\right)_P = \frac{8\pi T_H r_h^3}{8\pi P r_h^2 + \frac{3 Q^2}{r_h^2}-\left(1+c_2 m^2\right)}.
\end{eqnarray}
Condition for locally stable black hole is thus,
\begin{equation}
8\pi P r_h^2+\frac{3 Q^2}{r_h^2} > 1+c_2 m^2.
\end{equation}
In this work, we will use topological methods to understand the local and global stability of black holes in dRGT massive gravity theory. As mentioned before, the key idea is that a black hole solution is realized as thermodynamic defect. We discuss briefly the connection between topology and black hole solutions in the following section.

\section{Topology of Black Hole Thermodynamics}\label{topology}

In the topological approach, a black hole is described by generalized free energy defined by considering the black hole mass and temperature as independent parameters given by\cite{York:1986it},
\begin{equation}\label{general}
\mathcal{F} = M -\frac{S}{\tau}.
\end{equation}
Here, the parameter $\tau$ can be varied arbitrarily and has the dimension of time. For further analysis, we will be varying $\tau$ instead of black hole mass. Note that, when $1/\tau = T_H$, the black hole parameters satisfy the first law relation making the free energy on-shell, i.e., Eq. \ref{free}. \\

As mentioned in the introduction,  we construct a vector field $\phi$ as,
\begin{equation}\label{phi_field}
\phi = \left(\frac{\partial \mathcal{F}}{\partial r_h}, -\cot \Theta \ \csc \Theta\right),
\end{equation}
here, the variable  $0\leq \Theta \leq \pi$ is used for convenience \cite{Cunha:2020azh}. Note that the term $\phi^{\Theta} = -\cot \Theta \ \csc \Theta$ becomes infinite when $\Theta=0, \text{or} \ \pi$. As discussed before, the extremal points of the free energy corresponding to $\tau=1/{T_H}$ is obtained when $\Theta={\pi}/{2}$. At this point, the zero point of the component $\phi^{r_h}$ satisfies the black hole solution.
In Duan's $\phi-$ mapping topological current theory \cite{Duan:1979ucg, Duan:1984ws}, one can define a topological current as,
\begin{equation}
j^\mu = \frac{1}{2 \pi} \epsilon ^{\mu \nu \rho} \epsilon _{ab} \partial_\nu n^a \partial_\rho n^b; \qquad \mu, \ \nu, \ \rho =0,1,2,
\end{equation}
where $\partial_\nu=\frac{\partial}{\partial x^\nu}$ with $x^\nu=(\tau, r_h, \Theta)$. The normalized vector $n =(n^1, n^2)$ is defined as $n^a = \frac{\phi^a}{|| \phi||}; (a=1, \ 2)$, with $\phi^1=\phi^{r_h}$ and $\phi^2=\phi^\Theta$. Note that the topological current is conserved, i.e., $\partial_\mu j^\mu=0$. In order to calculate the topological number, we re-express the topological current as \cite{Wei:2020rbh},
\begin{equation}
j^\mu = \delta^2(\phi) J^\mu(\frac{\phi}{x}),
\end{equation}
where the Jacobi tensor is given by,
\begin{equation}
\epsilon^{ab}J^\mu(\frac{\phi}{x}) = \epsilon^{\mu \nu \rho} \partial_\nu \phi^a \partial_\rho \phi^b.
\end{equation}
Now, $j^\mu$ is non-zero only at $\phi=0$. The corresponding topological current density is given by,
\begin{equation}
j^0 = \sum_{i=1}^N \beta_i \eta_i \delta^2\left(\vec{x} -\vec{z}\right),
\end{equation}
where the Hopf index $\beta_i$ counts the number of the loops that the vector $\phi^\mu$ makes when $x^\mu$ circularize $z_i$. The coefficient $\eta_i$ is the Brouwer degree which satisfies $\eta_i = sign\left(J^0\left(\frac{\phi}{x}\right)_{z_i}\right) = \pm 1$, where $J^0\left(\frac{\phi}{x}\right) = \frac{\partial \left(\phi^1,\phi^2\right)}{\partial \left(x^1,x^2\right)}$. Finally, the topological number for a given parameter region $\Sigma$ is obtained as,
\begin{equation}
W = \int_{\Sigma} j^0 d^2 x = \sum_{i=1}^N \beta_i \eta_i = \sum_{i=1}^N w_i.
\end{equation}
Here, $w_i$ is the winding number corresponding to the $i^{\text{th}}$ zero point of $\phi$ contained in $\Sigma$. The value of the winding number is directly linked to the stability of black holes. A  positive winding number corresponds to a stable and a negative winding energy indicates an unstable black hole state. The topological number $W$ reflects the global topological features.

\section{Topology of Black Hole Thermodynamics in 4 Dimensions}\label{massive_topology}

In this section, we consider dRGT massive gravity theory in four dimensions. The black hole solutions described by the metric function Eq. \ref{metric_1}, exhibit small-large black hole phase transition for certain values of the parameters of the theory \cite{Fernando:2016qhq}. The previous studies on the thermodynamic characteristics of black holes in massive gravity theory were carried out by examining the free energy profile and the heat capacities using various techniques. Here, we employ the topological method to examine the local and global stability, and hence deduce the topological classification of these black hole solutions. First, we obtain an expression for generalized free energy of the system. Once we have the free energy, we can apply the method detailed in the previous section to establish a parameter space and find the zero points of the corresponding vector field in it. The topological number characterizing the on-shell black hole solutions are then calculated using Duan's $\phi-$ mapping topological current theory. We begin with the simple case in which the black hole charge is zero.
\subsection {When $Q=0$}
\begin{figure}[ht!]
\center
\includegraphics[width=8cm]{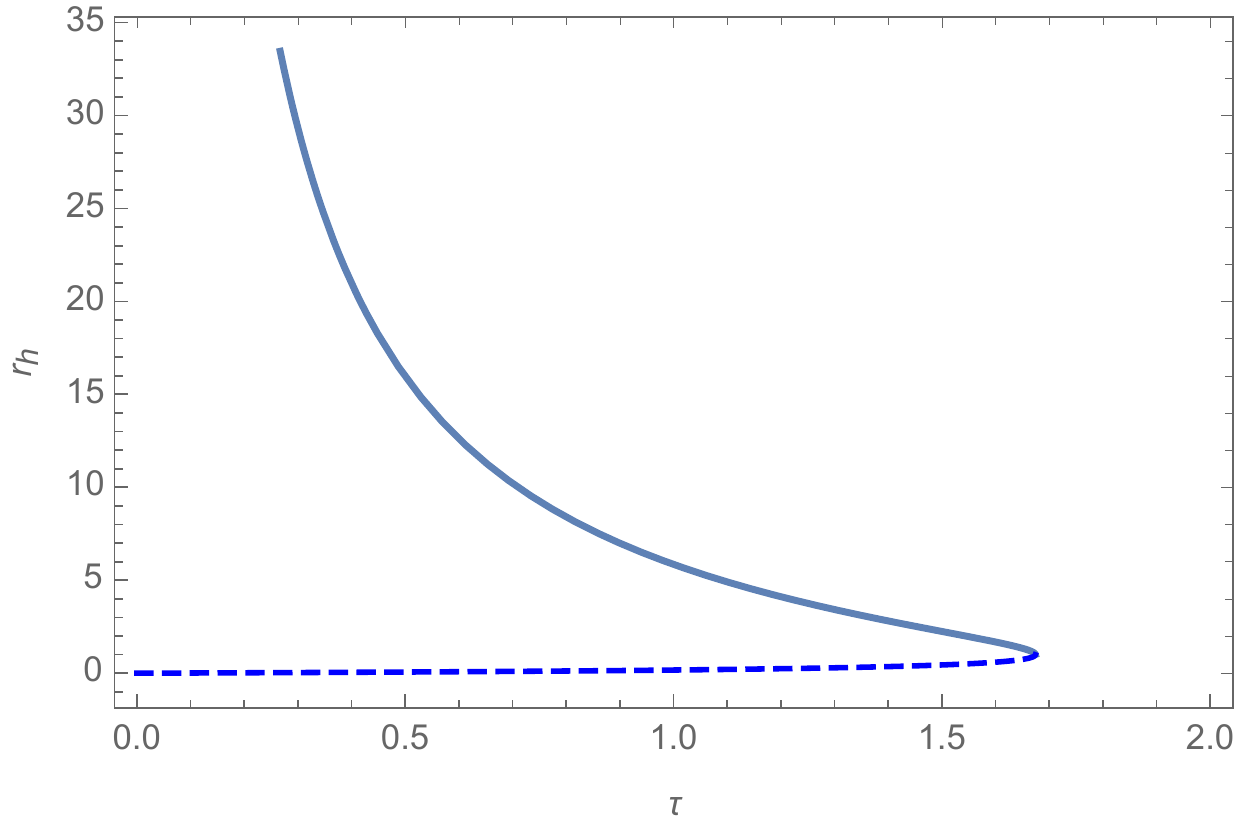}
\caption{\small{The curve between $\tau-r_h$ for $P=0.05, c_1=10, c_2=0.5,$ and $m=\sqrt{2}$. Solid and dashed lines represent the large and small black holes respectively. Asymptotically, $\tau \rightarrow 0$ for $r_h \rightarrow r_{min}$ and $r_h \rightarrow \infty$}}
\label{fig.p1}
\end{figure}

 The generalized free energy for the neutral black hole is obtained using Eq. \ref{general} as,
\begin{equation}\label{free_2}
\mathcal{F} = \frac{r_h}{2}\left(1+\frac{8\pi}{3} P r_h^2 + m^2(c_2+\frac{1}{2} c_1 r_h)\right) - \frac{\pi r_h^2}{\tau}.
\end{equation}
Now, the component $\phi^{r_h}$ is given by,
\begin{equation}
\phi^{r_h} = \frac{\partial \mathcal{F}}{\partial r_h} = \frac{1}{2} + 4 \pi P r_h^2 +\frac{m^2}{2}\left(c_2+c_1 r_h\right) - \frac{2 \pi r_h}{\tau}.
\end{equation}
Solving $\phi^{r_h}=0$, we obtain an expression for the parameter $\tau$ as,
\begin{equation}\label{tauzero}
\tau = \frac{4\pi r_h}{1+8 \pi P r_h^2 + {m^2}\left(c_2+c_1r_h\right)}.
\end{equation}
Note that the time parameter is fixed to be $\tau = \frac{1}{T_H}$ by specifying the ensemble temperature. Therefore, the solutions of Eq. \ref{tauzero} correspond to the on-shell black holes at some value of $T_H$. A plot of $\tau-r_h$ for a set of parameters of the uncharged black hole in massive gravity theory is shown in Fig. \ref{fig.p1}. One can see that there are two on-shell black holes namely small and large black holes, and no solutions for $\tau>\tau_c$. Further, the time parameter $\tau \rightarrow 0$ for $r_h \rightarrow r_{min}$ and $r_h \rightarrow \infty$. Here, $r_{min}$ is the minimum value of $r_h$ (zero for Schwarzschild metric). Interestingly, these results are similar to the case of spherically symmetric black holes in Einstein-Gauss-Bonnet theory for dimensions $D\geq 6$ \cite{Liu:2022aqt}. We analyse the stability of these black hole states by calculating Eq. \ref{free_2}. It is easy to show that the black hole with the large radius is globally stable (global minima on the free energy landscape) whereas the one with the small radius is locally unstable (local maximum on the free energy landscape). 

\begin{figure}[ht!]
\center
\includegraphics[width=8cm]{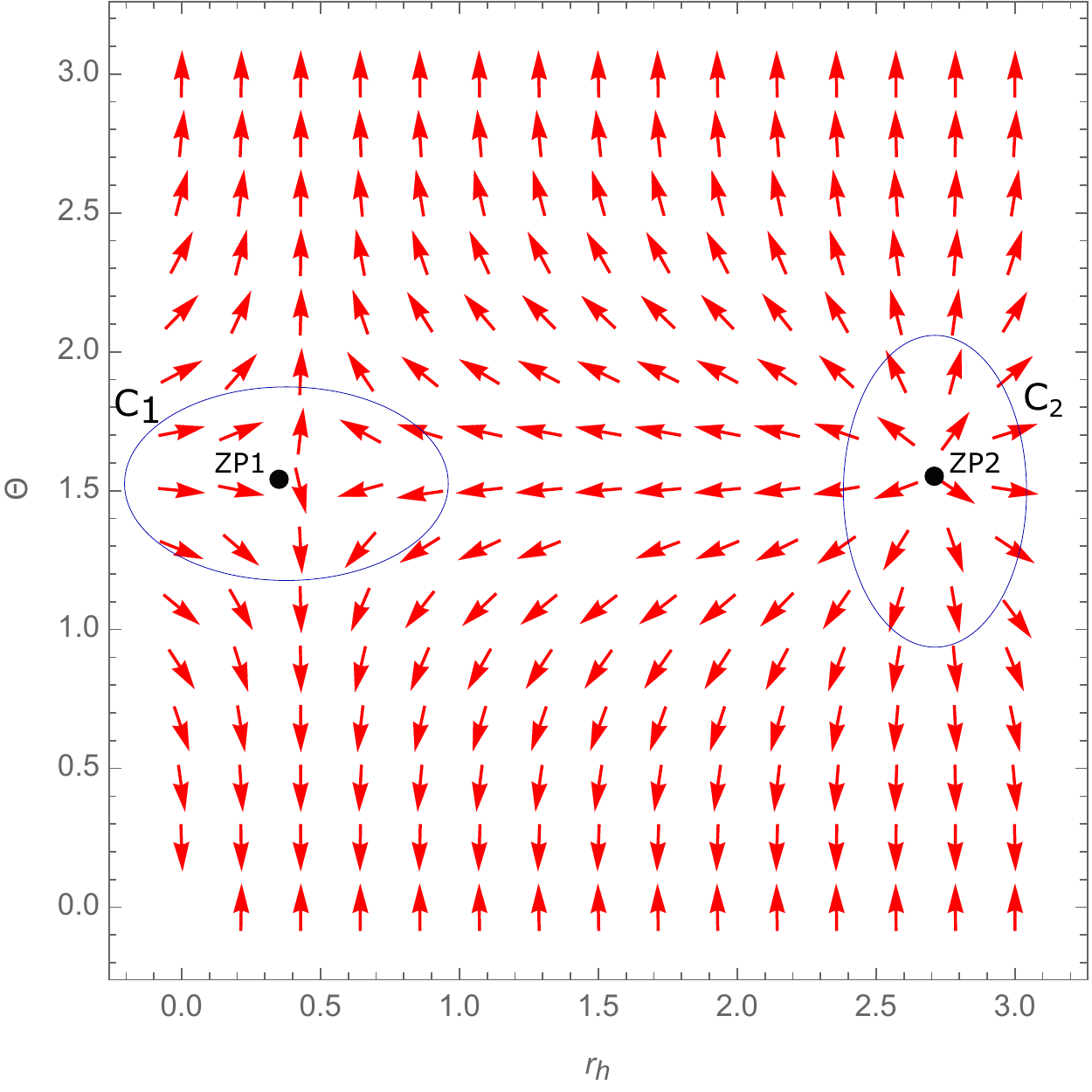}
\caption{\small{The arrows represent the unit vector $n$ on $r_h-\Theta$ plane for $\tau=0.9$. The black dots represent the zero points $ZP_1$ (small black hole) and $ZP_2$ (large black hole)}}
\label{fig.p2}
\end{figure}

Now, we plot the unit vector field $n$ for some arbitrarily chosen value of $\tau$ ( say, $\tau=0.9$) in Fig. \ref{fig.p2}. We find two zero points $ZP_1$ and $ZP_2$ corresponding  to small and large black hole states respectively. We calculate the winding number $w$ for some arbitrary loops $C_1$ and $C_2$. A simple calculation shows that the winding number for $C_1$ is $-1$, whereas for $C_2$ is $+1$. Further, the conjecture that the stable (unstable) black hole corresponds to positive (negative) value of winding number holds in the case of uncharged black hole solutions in dRGT massive gravity theory. As discussed before, the topological number characterizes the global properties. Moreover, the global properties are used to identify different topological classes of black hole solutions. Here, the topological number satisfies $W=1+(-1)=0$, similar to the case of Reissner-Nordstrom black holes \cite{{Wei:2022dzw}}. However, one must consider the behaviour of the vector field at $r_h=r_{min}$ and $\infty$. In the case of uncharged black holes in massive gravity, the topological number is the same as that of RN black hole. However,  the asymptotic behaviour of $r_h$ is different as  the time parameter $\tau \rightarrow 0$ for both $r_h \rightarrow r_{min}$ and $r_h \rightarrow \infty$ on the $r_h-\tau$ plane. In the case of RN black hole, $\tau \rightarrow \infty$ for both  $r_h \rightarrow r_{min}$ and $r_h \rightarrow \infty$ Therefore, these two solutions are not topologically equivalent. This observation is reported for the case of spherically symmetric black holes in Einstein-Gauss-Bonnet theory for dimensions $D\geq 6$ \cite{Liu:2022aqt}. Hence, one can conclude that the  neutral black holes in four-dimensional massive gravity and $D\geq6$ Gauss-Bonnet gravity theories share the same topological class.

\subsection {When  $ Q\neq 0$}
\begin{figure}[ht!]
\center
\includegraphics[width=8cm]{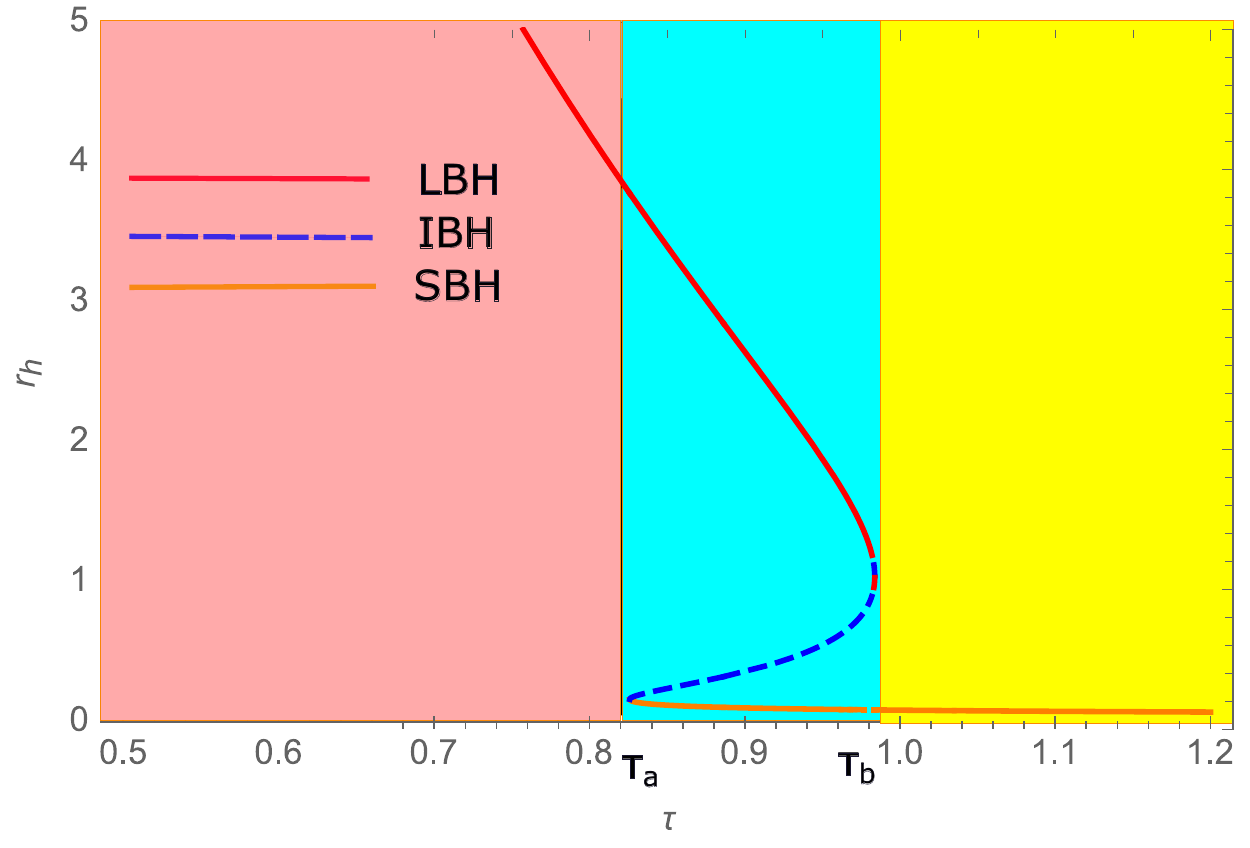}
\caption{\small{$r_h-\tau $ curve representing $\phi^{r_h} =0$ for $P=0.05, c_1=10, c_2=0.55$, and $m=\sqrt{2}$. Red solid, blue dashed, and orange solid lines represent large (LBH), intermediate (IBH), and small (SBH) black holes respectively. All three black holes exist for $\tau_a<\tau<\tau_b$. Irrespective of the number of black hole solutions, the topological number for different colour regions is same ($W=1$).}}
\label{charged_1}
\end{figure}

  Let us now consider a charged black hole in four-dimensional massive gravity theory. As detailed before, we start with the expression of generalized free energy, i.e., Eq. \ref{general}.
  \begin{eqnarray}
  \mathcal{F} = M-\frac{S}{\tau}= \frac{r_h}{2}\left( 1+8\pi P r_h^2 +\frac{Q^2}{r_h^2} + m^2\left(\frac{c_1}{2} r_h + c_2\right)\right)-\frac{\pi r_h^2}{\tau},
  \end{eqnarray}
 where $Q$ denotes the black hole charge. Note that the above expression for the free energy is off-shell, and the parameter $\tau$ is the Euclidean time period. The expression becomes on-shell when the time parameter is identified with the inverse of Hawking temperature. In the view point of topology, every black hole solution can be associated with a topological charge using a vector $\phi = (\phi^{r_h}, \phi^\Theta)$. Using Eq. \ref{phi_field},
  \begin{equation}
  \phi^{r_h} = \frac{\partial \mathcal{F}}{\partial r_h} = \frac{1}{2}\left(1+8 \pi P r_h^2 - \frac{Q^2}{r_h^2}+m^2\left(c_1 r_h+c_2\right)\right)- \frac{2\pi r_h}{\tau}.
  \end{equation}
  Now we solve $\phi^{r_h} = 0$ to obtain an expression for the parameter $\tau$ as,
  \begin{equation}\label{tauone}
\tau = \frac{4\pi r_h}{1+8 \pi P r_h^2 -\frac{Q^2}{r_h^2}+ m^2\left(c_2+c_1r_h\right)}.
\end{equation}
In Fig. \ref{charged_1}, we have plotted $r_h-\tau$ curve for Eq. \ref{tauone}. For certain values of the parameters, there exist three branches of solutions for $\tau_a < \tau < \tau_b$, where $\tau_a, \tau_b$ are some values of $\tau$ at which the phase transition occur.  We label them as small, intermediate, and large black holes. There exists only one black hole solution for  $\tau < \tau_a$ and $\tau> \tau_b$, corresponding to small and large black holes respectively. One can study the asymptotic behaviour of the vector by looking at profile of $\tau$. The parameter $\tau \rightarrow 0$ as $r_h \rightarrow \infty$ and $\tau \rightarrow \infty$ as $r_h \rightarrow r_{min}$.

\begin{figure}[ht!]
\center
\includegraphics[width=8cm]{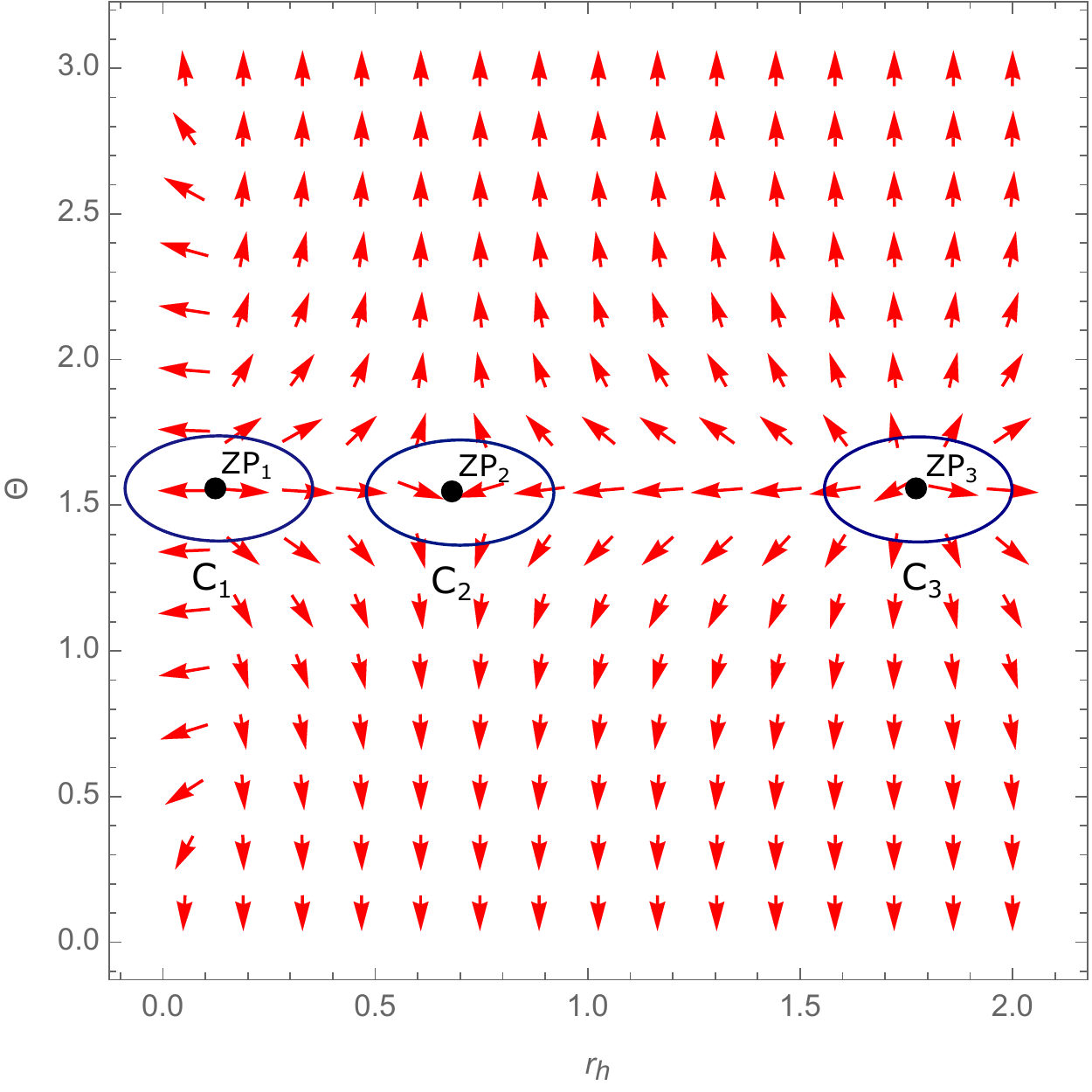}
\caption{\small{The unit vector on a portion of $r_h-\Theta$ plane for $P=0.05, c_1=10, c_2=0.55, m=1$, and $\tau=0.96$. The black dots represent the zero points (ZPs). The $z_i$s are enclosed by arbitrary contours $C_i$s. The winding number for $C_1, C_2$, and $C_3$ are $+1, -1$, and $+1$ respectively.}}
\label{charged_2}
\end{figure}

To obtain the topology, we plot the unit vector $n$ constructed from $\phi$ from some $\tau \in (\tau_a, \tau_b)$. We observe three zero points namely $ZP_1, ZP_2$, and $ZP_3$ corresponding to small, intermediate, and large black holes in Fig. \ref{charged_2}. To find the winding number we enclose $ZP_1, ZP_2$, and $ZP_3$ using some arbitrary loops $C_1, C_2$, and $C_3$ respectively. Consequently the winding number for $C_1, C_2$, and $C_3$ are $+1, -1,$ and $+1$ respectively. This shows that the small and large black holes are thermodynamically stable, whereas the intermediate solution is unstable. The topological number is obtained as $W=1-1+1=1$, which does not depend on the value of $Q$. This behaviour of black hole phase transition in four-dimensional dRGT massive theory is same as in the case of  Reissner-Nordstrom anti-de Sitter (RN-AdS) black hole \cite{{Wei:2022dzw}}. Therefore, we report that the graviton mass does not alter the topological number in four dimensions. 

\section{Topology of Black Hole Thermodynamics in Higher Dimensions}\label{massive_topology_2}

\begin{figure}[ht!]
\center
\includegraphics[width=8cm]{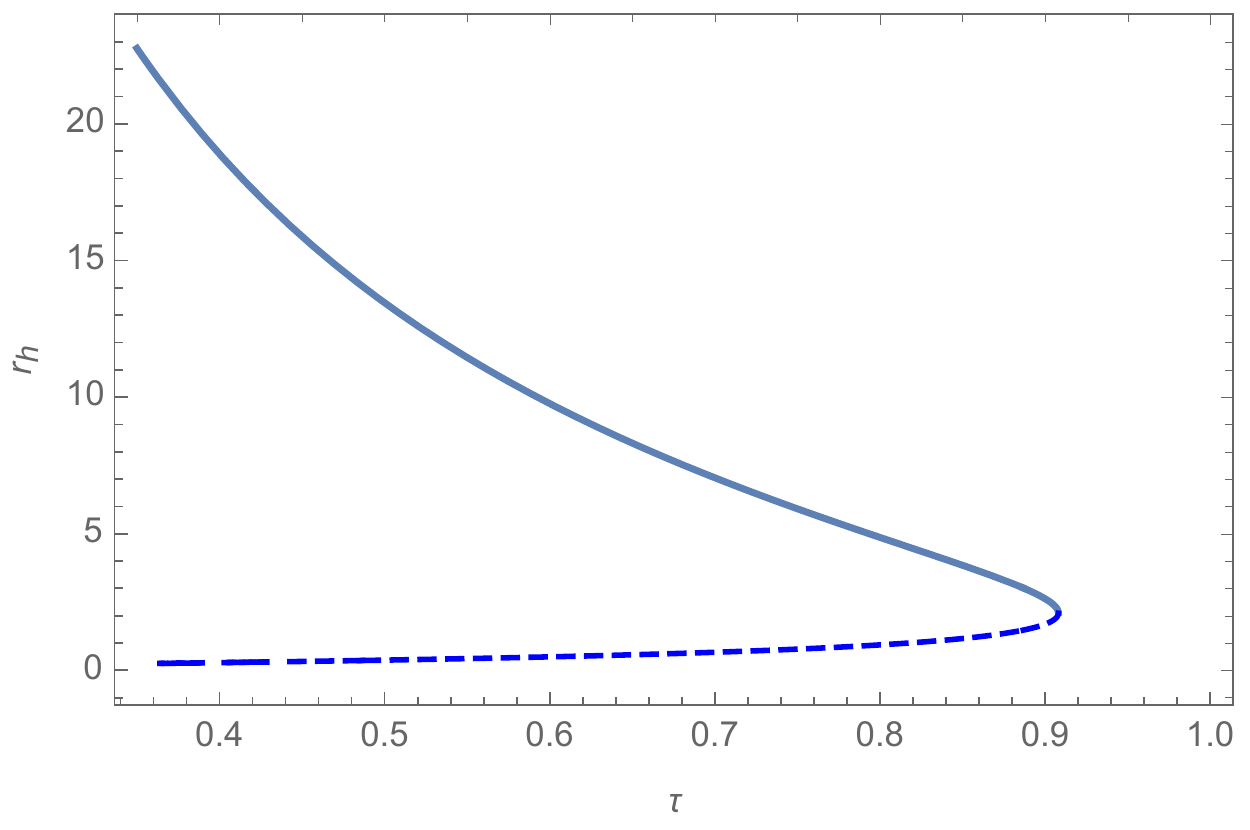}
\caption{\small{The curve between $\tau-r_h$ for $P=0.07, c_1=9, c_2=1.4, c_3=0.2$ and $m=1$. Solid and dashed lines represent the large and small black holes respectively. Asymptotically, $\tau \rightarrow 0$ when $r_h \rightarrow r_{min}$ and $\infty$.}}
\label{fig.p5}
\end{figure}

In this section, we discuss the topology classes of higher dimensional black hole solutions in dRGT massive gravity. The dimensional dependences of the topological number have been reported for various black hole solutions (see \cite{Bai:2022klw, Liu:2022aqt, Wu:2022whe}). Naturally, it is worth studying the effect of dimension in the case of massive gravity. To this extend, one can safely put $Q=0$. In five dimensions ($n=3$), one has $c_3\ne0$. Accordingly, the expression for the generalized free energy obtained from Eq. (\ref{m}, \ref{s}) is given by,

\begin{eqnarray}
\mathcal{F} = M - \frac{S}{\tau} = \frac{3 \pi}{8}r_h \Big[ \left(1+c_2 m^2\right) r_h+ \frac{4 \pi}{3} P r_h^3 + \frac{1}{3}c_1 m^2 r_h^2 + 2 c_3 m^2\Big] - \frac{\pi^2}{2 \tau} r_h^3.
\end{eqnarray}
Following the construction developed in \ref{topology}, the component $\phi^{r_h}$ of the vector field is obtained as,
\begin{equation}
\phi^{r_h} = \frac{\partial \mathcal{F}}{\partial r_h} = \frac{3\pi}{4} \Big[ \left(1+c_2 m^2\right) r_h + \frac{2 \pi P}{3} r_h^3 + \frac{1}{2} c_1 m^2 r_h^2 + c_3 m^2\Big] - \frac{3 \pi^2}{2\tau} r_h^2.
\end{equation}
Now, the parameter $\tau$ is obtained by solving $\phi^{r_h}=0$,
\begin{equation}
\tau = \frac{2 \pi r_h^2}{\left(1+c_2m^2\right) r_h+c_3m^2+\frac{1}{2}c_1 m^2 r_h^2+\frac{8 \pi}{3} P r_h^3}.
\end{equation}
We observe that the sign of the parameter $c_3$ alters the topology of neutral black holes in five-dimensional massive gravity. Therefore, we analyse $c_3>0$ and $c_3<0$ separately. In the case where $c_3>0$, the results are qualitatively similar to the neutral black hole in four-dimensional massive gravity. In Fig. \ref{fig.p5}, a plot of $\tau-r_h$ is given for $c_3>0$. The plot shows the Hawking-Page phase transition between small and large black holes at some $\tau=\tau_c$, and there exist two black holes for $\tau<\tau_c$ and no solutions for $\tau>\tau_c$. The black hole with the small radius is thermodynamically unstable (negative heat capacity) whereas, the large black hole is stable. Also, the parameter $\tau \rightarrow 0$ as $r_h \rightarrow r_{min}$ and $r_h \rightarrow \infty$.

\begin{figure}[ht!]
\center
\includegraphics[width=8cm]{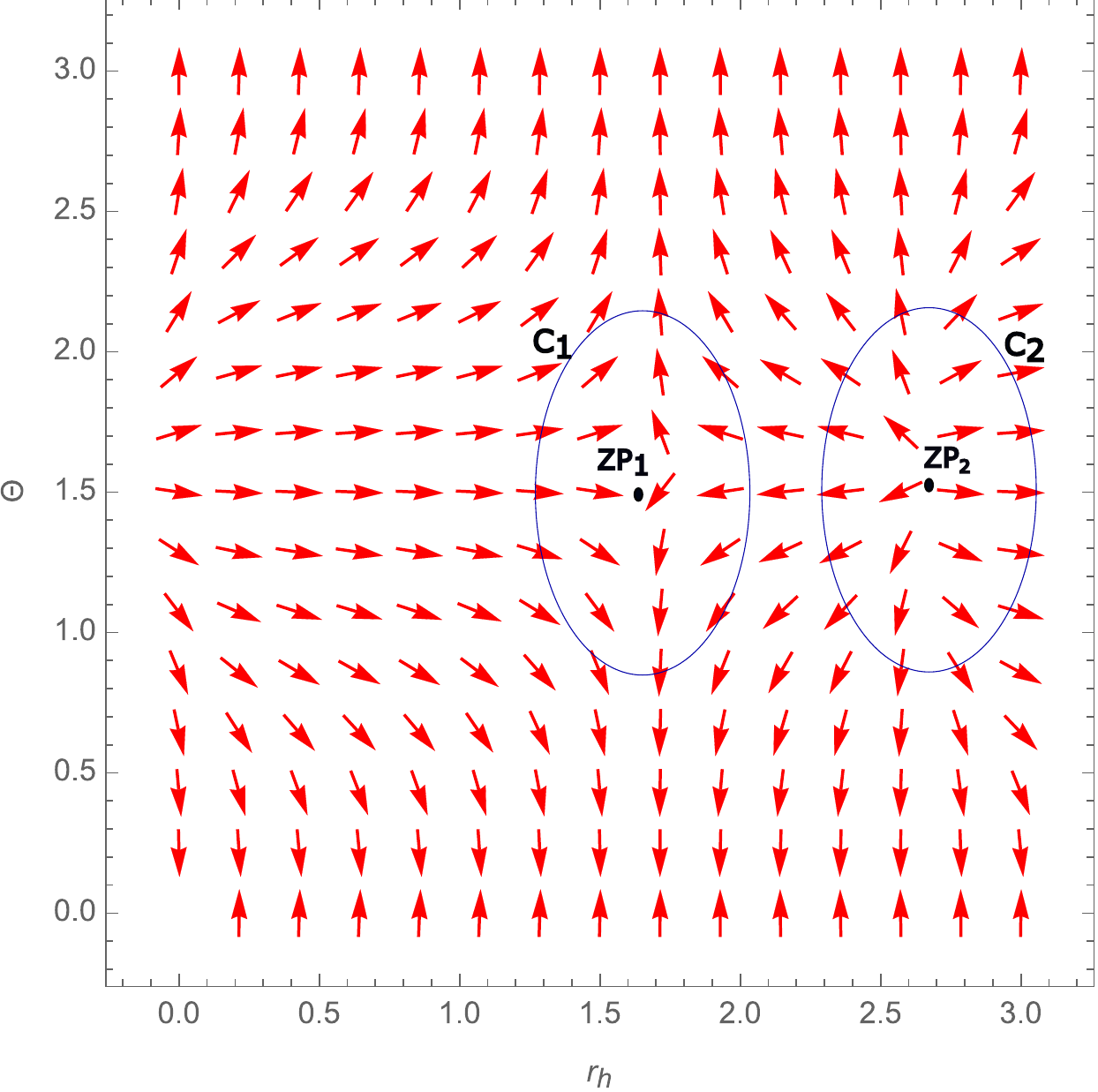}
\caption{\small{The arrows represent the unit vector $n$ on $r_h-\Theta$ plane for $\tau=0.9$. The black dots represent the zero points $ZP_1$ (small black hole) and $ZP_2$ (large black hole)}}
\label{fig.p6}
\end{figure}

A plot of unit vector field $n$ is given in Fig. \ref{fig.p6}. The zero points $ZP_1$, $ZP_2$ represent small and large black hole phases. The winding number corresponding to $ZP_1$ and $ZP_2$ are $w=-1$ and $w=+1$ respectively. Further, the topological number $W=1+(-1)=0$. Considering the topological number and the asymptotic behaviour of $r_h$, we conclude that the topology class of neutral black holes in five-dimensional massive gravity with positive $c_3$ is the same as of the neutral black hole in four-dimensional massive gravity and the Einstein-Gauss-Bonnet black holes in dimensions $D\geq6$.

\begin{figure}[ht!]
\center
\includegraphics[width=8cm]{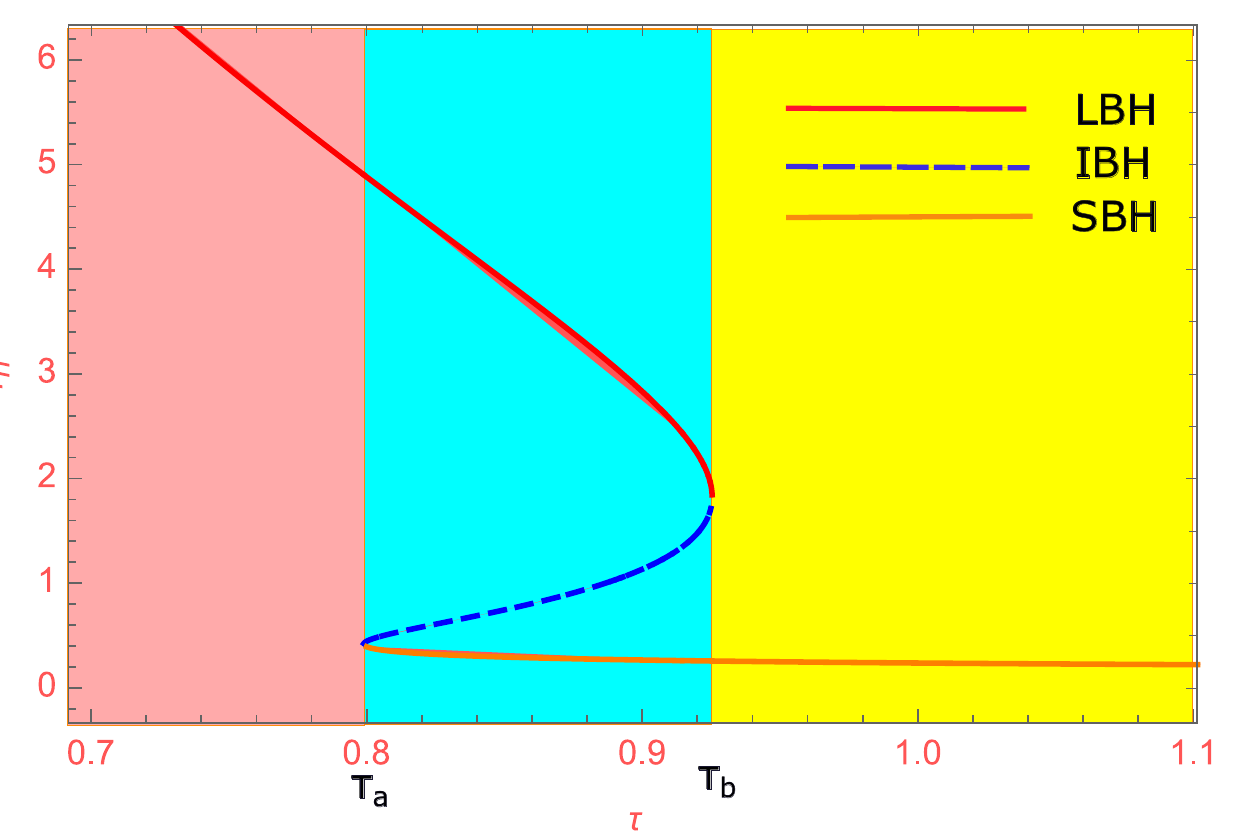}
\caption{\small{The curve between $\tau-r_h$ for $P=0.07, c_1=9, c_2=1.5, c_3 =-0.5$, and $m=1$. Solid and dashed lines represent the large and small black holes respectively. Asymptotically, $\tau \rightarrow 0$ when $r_h \rightarrow r_{min}$ and $\infty$.}}
\label{fig.p7}
\end{figure}

Interesting results are obtained when we put $c_3<0$. We observe that the term $c_3 m^2$ plays the role of charge. The $\tau-r_h$  plot shown in Fig. \ref{fig.p7} is qualitatively same as of charged black holes in four-dimensional massive gravity Fig. \ref{charged_1}. There exist three solutions in the range $\tau_a < \tau < \tau_b$, namely, large, intermediate, and small black hole phases. There exist only one solution for both $\tau<\tau_a$ (large BH) and $\tau>\tau_b$ (small BH). Here, both small and large black holes are thermodynamically stable whereas the intermediate one is unstable. The asymptotic behaviour of $r_h$ can be easily deduced. The parameter $\tau \rightarrow 0$ as $r_h \rightarrow \infty$ and $\tau \rightarrow \infty$ as $r_h \rightarrow r_{min}$.

\begin{figure}[ht!]
\center
\includegraphics[width=8cm]{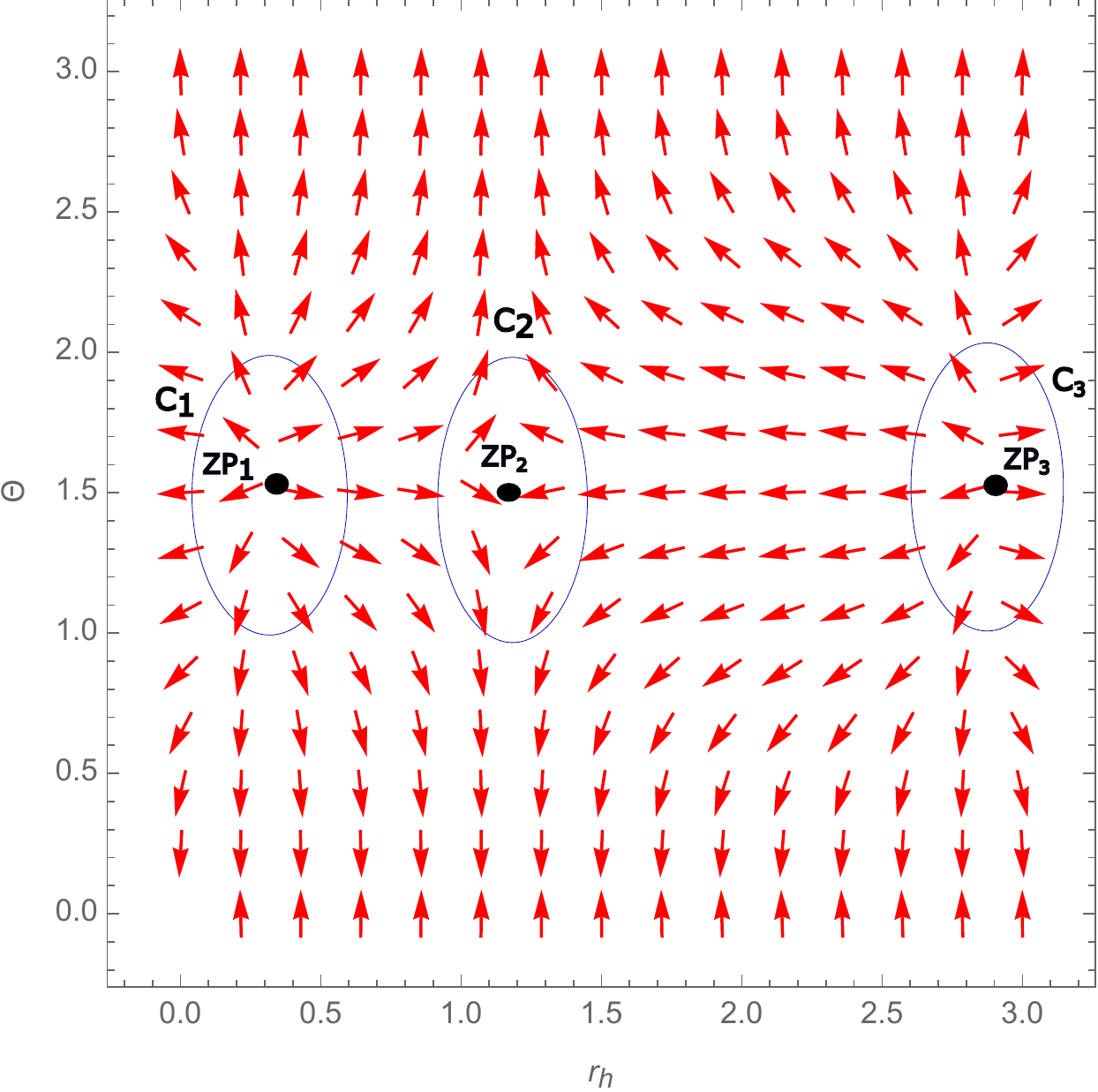}
\caption{\small{The unit vector on a portion of $r_h-\Theta$ plane for $P=0.05, c_1=10, c_2=0.55, m=1$, and $\tau=0.96$. The black dots represent the zero points (ZPs). The blue contours $C_i$ are arbitrary closed loops enclosing the zero points. The winding number for $C_1, C_2$, and $C_3$ are $+1, -1$, and $+1$ respectively.}}
\label{fig_fin}
\end{figure}

The topological number is obtained by plotting the unit vector field $n$ constructed from $\phi$ on $\Theta-r_h$ plane Fig. \ref{fig_fin}. The three solutions namely, small, intermediate, and large black hole phases are described by the zero points $ZP_1, ZP_2$ and $ZP_3$ respectively. The winding number corresponding to these zero points are $+1, -1,$ and $+1$ respectively. The topological number is thus obtained as $W= 1-1+1=1$. Analysing the topological number as well as the asymptotic behaviour of $\tau$ with respect to the parameter $r_h$, we conclude that the topology class of neutral black holes in five-dimensional massive gravity with $c_3<0$ is same as of charged solution in four-dimensional massive gravity and RN- AdS black holes.\\

\section{discussion}\label{dis}
In this paper, we have analysed the local and global thermodynamic stability conditions of black holes in dRGT massive gravity theory using Duan's $\phi-$mapping topological current theory. The topological number which is used to group black holes into different topological classes, is estimated by treating the black hole solutions as thermodynamic defects. A vector field is constructed from the generalized free energy, where zero points of the vector correspond to on-shell black hole solutions. Further, we have calculated the winding number associated with each zero point which reflects the local thermodynamic stability. A positive (negative) winding number corresponds locally thermodynamically stable (unstable) black hole solution. \\

We have analysed the topological nature of both charged and uncharged  black holes in dRGT massive gravity. In four dimensions with black hole charge $Q=0$, the vector field $\phi$ (Eq. \ref{phi_field}) admits two zero points representing small (SBH) and large (SBH) black hole solutions. The winding number for SBH is $w_1=-1$ and for LBH is $w_2=+1$. Therefore, the topological number $W =w_1+w_2 = 0$. Even though this topological number is the same as that of RN black holes, the asymptotic behaviour of $\tau$ as $r_h \rightarrow r_{min}$ and $r_h \rightarrow \infty$ are different. However, the topological number as well as the asymptotic behaviour of $\tau$ of neutral black holes in massive gravity are the same as  that of spherically symmetric black holes in Einstein-Gauss-Bonnet theory for dimensions $D\geq 6$ \cite{Liu:2022aqt}. Therefore, these two solutions share the same topology class.\\

In the case when $Q\neq 0$, the vector $\phi$ admits three zero points, corresponding to small, intermediate, and large black holes. The topological number is estimated to be $W=+1$. The value of $W$ and the asymptotic behaviour of $\tau$ are the same as that of AdS-RN black holes. This suggests that the graviton mass does not alter the topological number in four dimensions. Therefore, our result strengthens the proposition that the modification of GR does not alter the topological characteristics.\\

In higher dimensions, our result rules out the conjecture that the topological number at an arbitrary temperature is a universal independent of black hole parameters \cite{Wei:2022dzw}. The sign of dRGT massive gravity parameter $c_3$, determines the topology number as well as the asymptotic behaviour of $\tau$.  In five dimensions, black hole solution with $c_3>0$ carries a topological number $W=0$ and the solutions with $c_3<0$ carries a topological number $W=0$.\\

It has been argued that the black holes in GR, as well as modified gravity theories, can be classified based on their topological nature. Accordingly, topological number $W= -1, 0, +1$ denote three distinct topological classes \cite{Wei:2022dzw,Wu:2022whe, Liu:2022aqt, Bai:2022klw}. Further, one can identify the topologically equivalent black hole solutions by examining the asymptotic behavior of the parameter $\tau(r_h)$. Therefore, AdS-RN, charged black holes in massive gravity, neutral black holes in five-dimensional massive gravity with $c_3<0$, and charged Gauss-Bonnet black holes in AdS spacetime belong to the same topology class with $W=+1$. Whereas, black holes in Einstein-Gauss-Bonnet theory for dimensions $D\geq6$, neutral black holes in four-dimensional massive gravity, and neutral black holes in five-dimensional massive gravity with $c_3>0$ share the same topology class with topological number $W=0$. The results we have presented are valid for black holes with spherical horizon hypersurfaces. One can revisit the calculations to understand the topological nature of black hole solutions for planar and hyperbolic horizon hypersurfaces.\\

\section{Acknowledgements}
TS thanks Balasundaram Muthukumar for guidance and support.
\appendix

\end{document}